\documentclass[aps,pra,showpacs,twocolumn]{revtex4-1}
\usepackage{amssymb}
\usepackage{amsmath}
\usepackage{graphicx}
\usepackage{epsfig}
\usepackage{subfigure}
\usepackage{color}

\begin{document}

\title{Dynamic generation of nonequilibrium superconducting states in
Kitaev chain}
\author{Y. B. Shi, K. L. Zhang}
\author{Z. Song}
\email{songtc@nankai.edu.cn}

\begin{abstract}
A non-equilibrium state shares macroscopic properties, such as conductivity
and superconductivity, with a static state when the two states exhibit an
identical average value of an observable over a period. We studied the
quench dynamics of a Kitaev chain on the basis of two types of order
parameters associated with two channels of pairing: local pairing in real
space and Bardeen-Cooper-Schrieffer (BCS)-like pairing in momentum space. On
the basis of the exact solution, we found that the two order parameters were
identical for the ground state, which indicated a balance between the two
pairing channels and would help determine the quantum phase diagram.
However, for a non-equilibrium state obtained through time evolution from an
initially prepared vacuum state, the two parameters varied; however, both
parameters could still help determine the phase diagram. In the region of
nontrivial topological phases, non-equilibrium states favored BCS-like
pairing. This paper presents an alternative approach to dynamically generate
a superconducting state from a trivial empty state and illuminates the
pairing mechanism.
\end{abstract}

\maketitle
\affiliation{School of Physics, Nankai University, Tianjin 300071, China}

\section{\qquad Introduction}

The topic of non-equilibrium phenomena in quantum many-body systems is
fundamental and compelling in condensed matter physics. In conventional
theory, quantum phase transitions occur at zero temperature, and a phase
diagram refers, in particular, to the ground state of a system. However, it is obviously that the ground state is intimately related to the
excited states for a certain class of systems since they all stem from a
same time-independent Hamiltonian. For instance, for a Hamiltonian written
in pseudospin form $\sum_{k}\mathbf{B}_{k}\mathbf{\cdot s}_{k}$, the ground
state is determined by the field distribution $\mathbf{B}_{k}/\left\vert 
\mathbf{B}_{k}\right\vert $ \cite{ZG}, which can be extracted from any
eigenstates in the tensor product form. In this sense, non-equilibrium
phenomena within the prethermalization regime can probably reflect the phase
diagram from an alternative aspect. A well-known probe of many-body quantum
dynamics is quantum quench, in which a many-body system is initially
prepared in a state that is typically the ground state of some simple
Hamiltonian and subsequently evolves with a different Hamiltonian. In a
conventional framework, the properties of prequench and postquench
Hamiltonians are extracted from the evolved states instead of the ground
state. The use of nonequilibrium many-body dynamics presents an alternative
approach for accessing a new exotic quantum state with an energy level
considerably different from that of the ground state \cite%
{Choi,Else,Khemani,Lindner,Kaneko,Tindall,YXMPRA,ZXZPRB2,TK,JT1,JT2}.
Advancements in atomic physics, quantum optics, and nanoscience have allowed
the development of artificial systems with high accuracy \cite{Jochim,
Greiner}. Thus, theoretical models can be simulated and realized
experimentally. Direct simulation of a simple model is useful not only for
solving key challenges in condensed matter physics but also in the
engineering design of quantum devices. Notably, the availability of an
experimentally controllable fermion system provides an unprecedented
opportunity to explore nonequilibrium dynamics in interacting many-body
systems.

In addition, it has recently been established that experiments
with various materials and excitation conditions have witnessed phenomena
with no equilibrium analog or accessibility of chemical substitution,
including superconducting-like phases \cite{AC,DF,MM,TS}, charge density
waves (CDW) \cite{LS,HM,AK} and excitonic condensation \cite{YM}. Therefore,
how to stabilize a system in a non-equilibrium superconducting phase with a
long lifetime is a great challenge and is at the forefront of current
research. Among various non-equilibrium protocols, the generation of the $%
\eta $-pairing-like state plays a pivotal role in which the existence of
doublon and holes facilitate the superconductivity \cite%
{SK,TK,JT,FP,RF1,RF2,SE,JL,TK2,ZXZPRB2,ZXZPRB3,JT1,JT2}.

In the present paper, we report our theoretical investigation of
nonequilibrium dynamics in a paradigmatic quantum many-body model---the
Kitaev chain. Comparing to the previous works on the Hubbard
model, the intial state is an easily prepared state \cite{YXMPRA} and
quenched Hamiltonian is time-independent with parameters covering rich phase
diagram in the present work. The Kitaev chain is a lattice model of a
p-wave superconducting wire, which realizes Majorana zero modes at the ends
of the chain \cite{Kitaev}. Unpaired Majorana modes have been exponentially
localized at the ends of open Kitaev chains \cite{Sarma,Stern,Alicea}. The
primary feature of this model originates from the pairing term, which
violates the conservation of the fermion number but preserves its parity,
leading to a superconducting state. In general, most studies on this model
have centered on the ground state or the dynamical quantum
phase transitions related to the ground state \cite{MH,LZ}. The strength of
the pairing interactions has an important role in giving rise to a gapped
superconducting state. However, knowledge regarding the pair in each quantum
phase remains limited. Specifically, the nearest-neighbor (NN) pair creation
(annihilation) term is subtle for the formation of a pair. It results in an
NN pair in real space. By contrast, the Hamiltonian in k-space indicates
that it also contributes to Bardeen-Cooper-Schrieffer (BCS)-like pairing in
momentum space, which involves two particles with opposite momentum. In
addition, distinct features originate from the pair term in not only statics
but also dynamics. To determine the features of a given state, we introduced
two types of order parameters associated with two channels of pairing: local
pairing in real space and BCS-like pairing in momentum space. On the basis
of the exact solution, we found that the two order parameters were identical
for the ground state and could help determine the quantum phase diagram.
This revealed an unprecedented feature of the ground state that helps
balance between two types of pairing channels. The primary aim of this study
was to obtain pertinent information regarding the nonequilibrium state. In
this study, acquisition of a vacuum state by the quench process, in which a
prequench Hamiltonian has infinite chemical potential, was considered to be
the ground state. We investigated evolved states under the postquench
Hamiltonian with parameters covering the entire region and found two types
of order parameters that were different but could help determine the quantum
phase diagram. Our results yield two key findings: a superconducting state
can be dynamically generated from a trivial state, and the difference
between two types of order parameters for a nonequilibrium state indicates
the generation of various superconducting states from the ground state, thus
expanding the variety of the state of matter that is associated with
different channels of pairing.

This paper is organized as follows. In Sec. \ref{Model and
order parameters}, we present the model and introduce two types of order
parameters to investigate the feature of superconducting state. In Subsecs. %
\ref{K space} and \ref{R space}, we calculate the order parameters in $k$\
space and real space for the ground state, respectively. In Sec. \ref%
{Dynamics}, we investigate the features of dynamic behavior of the system.
Finally, we give a summary and discussion in Section \ref{sec_summary}. Some
details of the derivations are given in Appendix.

\begin{figure}[tbh]
\centering \includegraphics[width=0.5\textwidth]{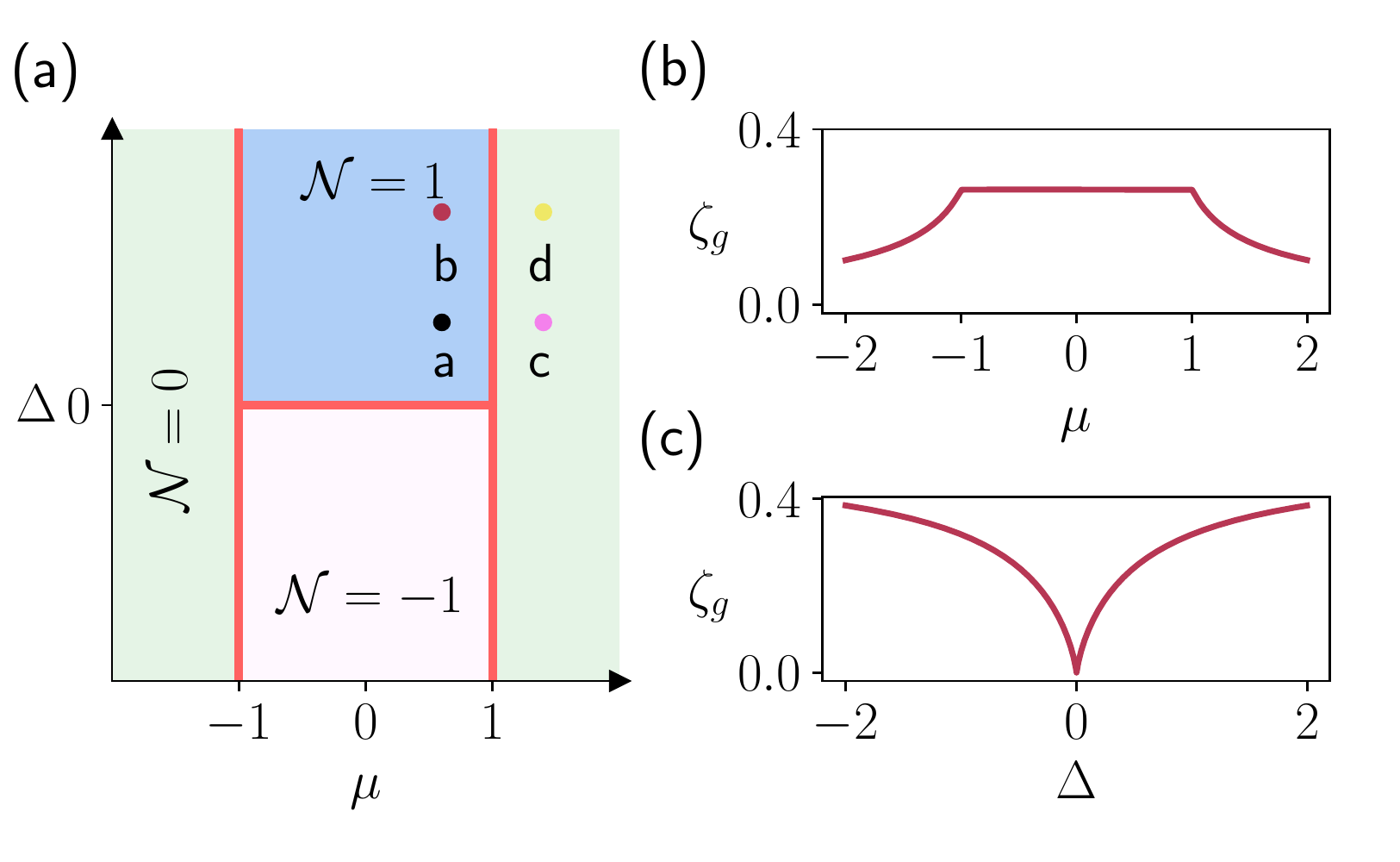} 
\caption{(a) Phase diagram of the Hamiltonian in Eq. (\protect\ref{H}) on
the $\protect\mu -\Delta $ plane. The sky blue, green and pink regions
correspond to the winding numbers $1,$ $0$ and $-1$, respectively. Red lines
indicate phase-transition lines. The four colored dots a-d represent the
typical sets of parameters at which the order parameters, as shown in\ Fig. 
\protect\ref{fig3}. (b) and (c), present the profiles of $\protect\zeta _{g}$\ defined in Eq. (\protect\ref{zeta_g}) at the lines $\Delta =0.6$ and $\protect\mu =0.6$, respectively. They indicate that $\protect\zeta _{g}$\ is
independent of $\protect\mu $\ in topologically non-trivial regions\ and is
non-analytic at the phase boundary. The other parameters are $N=2000$ and $J=1$.}
\label{fig1}
\end{figure}

\begin{figure*}[t]
\centering \includegraphics[width=1\textwidth]{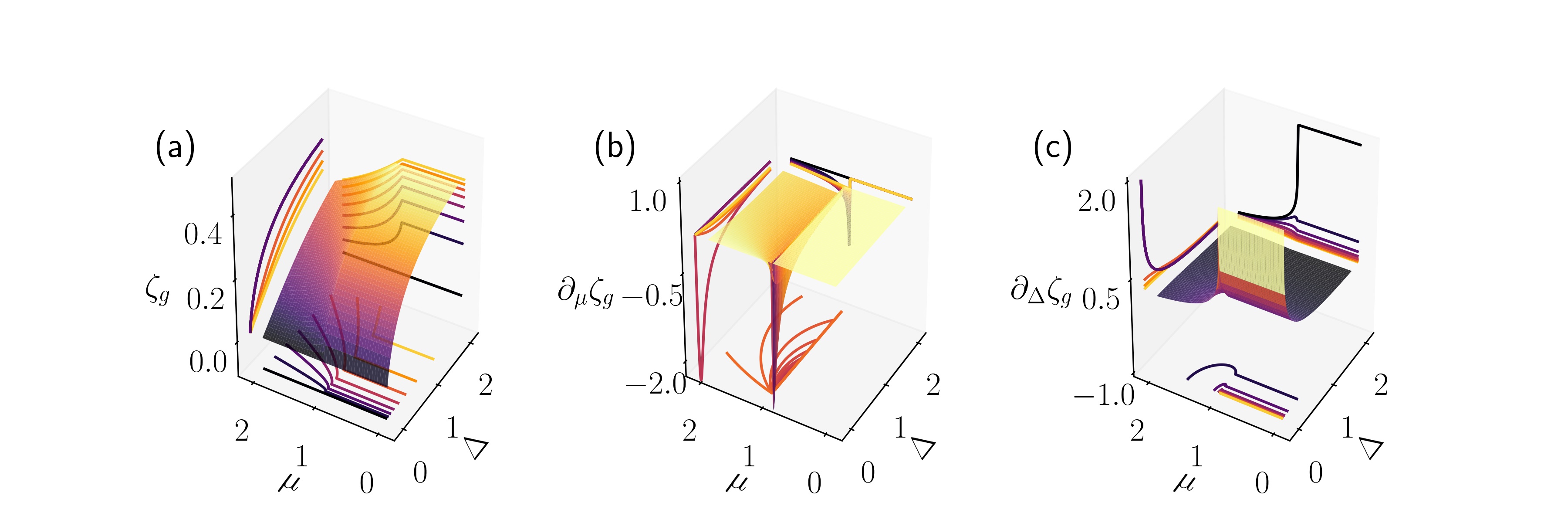}
\caption{Three-dimensional plots of the (a) order parameter of the ground
state $\protect\zeta _{g}$ obtained from Eq. (\protect\ref{zeta_g}); (b)
partial derivative of $\protect\zeta _{g}$ with respect to $\protect\mu ,$ $%
\partial _{\protect\mu }\protect\zeta _{g};$ and (c) partial derivative of $%
\protect\zeta _{g}$ with respect to $\Delta ,$ $\partial _{\Delta }\protect%
\zeta _{g},$ as the functions of $\protect\mu $ and $\Delta $ for case $%
N=4000$. The results indicate that second-order quantum phase transitions
occur at phase boundaries. Therefore, the analytical behavior of the order
parameter can identify the phase diagram. The other parameter is $J=1$. }
\label{fig2}
\end{figure*}

\begin{figure*}[tbh]
\centering \includegraphics[width=0.9\textwidth]{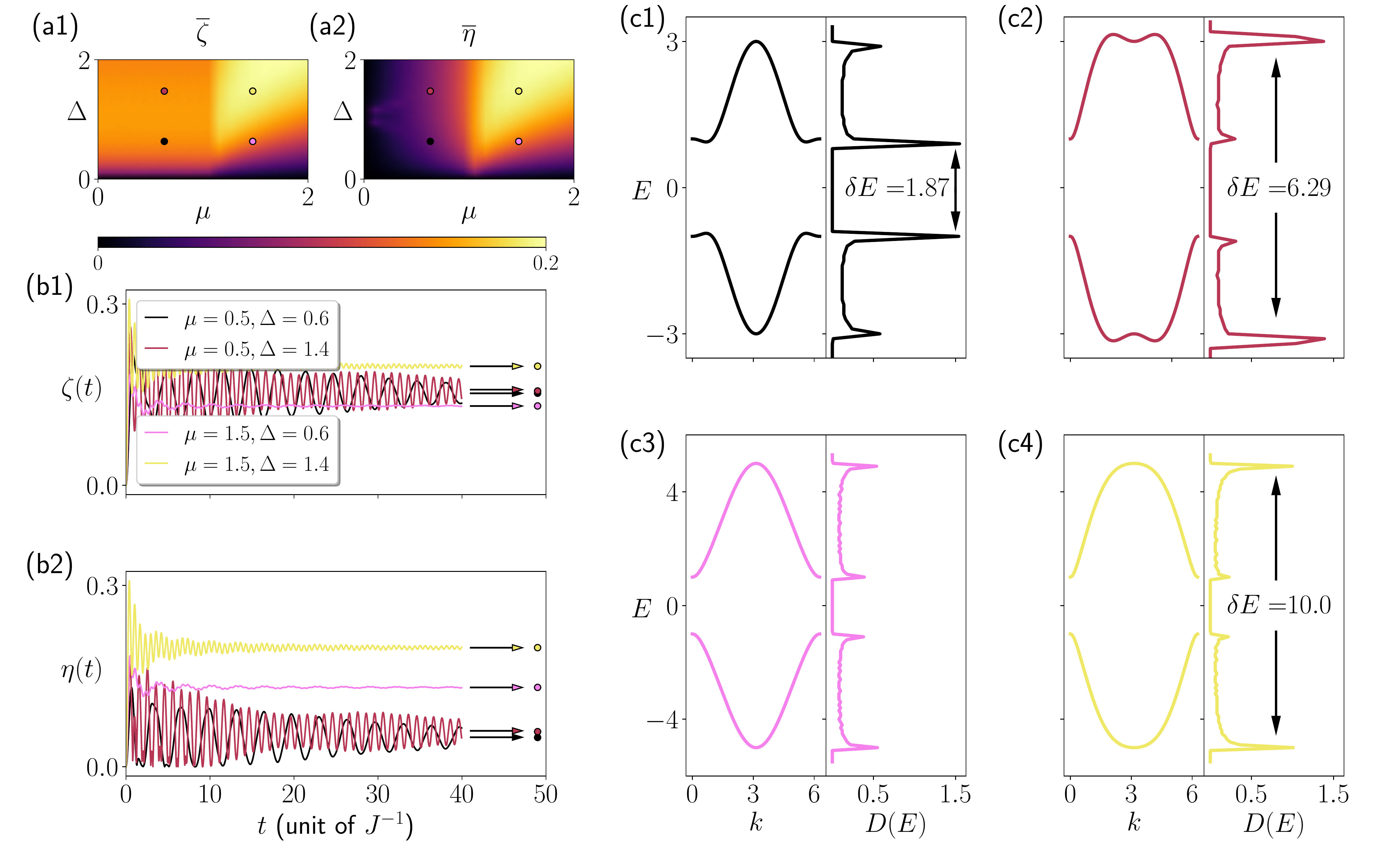}
\caption{(a) plots of the
		time-averaged analytical expressions for (1) $\zeta$ and (2) $\eta$ obtained from Eqs. (\ref{zeta}) and (\ref{eta}). The four colored dots
represent the dots a-d in Fig. \protect\ref{fig1}(a). (b) Profiles of (1) $%
\protect\zeta (t)$ and (2) $\protect\eta (t)$ obtained from Eqs. (\protect
\ref{etaT}) and (\protect\ref{zetaT})\ as the function of time for case $%
N=4000$ with four sets of parameters $(\protect\mu $ and $\Delta )$
indicated by dots a-d as shown in Fig. \protect\ref{fig1}(a). The colored
arrows are the stable values $\protect\zeta (t)$ and $\protect\eta (t)$ tend
to. The colored dots represent $\overline{\protect\zeta }$ and  $\overline{\protect\eta }$ obtained from Eqs. (\protect\ref{zeta}) and (%
\protect\ref{eta}) (c) Plots of dispersion $E=\pm \protect\varepsilon %
_{k}$ and DOS $D(E)$ for the systems with the same sets of parameters in
(b1) and (b2) with the same colors. We can see that both $\protect\zeta (t)$
and $\protect\eta (t)$ exhibit damping oscillations around their stable
values. The frequency and decay rate of amplitude are associated with the
energy difference $\protect\delta E$ between the highest peaks and maximum
of DOS. The amplitude decays slower if DOS has a higher peak. A larger value
of $\protect\delta E$ indicates a higher oscillation frequency. The other
parameter is $J=1$.}
\label{fig3}
\end{figure*}

\begin{figure*}[tbh]
\centering
\includegraphics[width=1\textwidth]{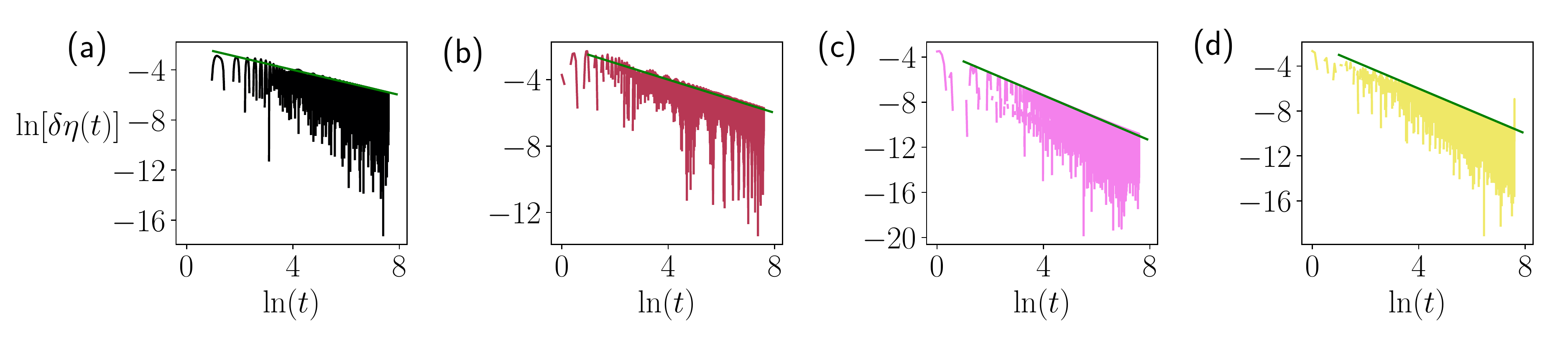}
\caption{Plots of $\ln [\protect\delta \protect\eta (t)]$ as
functions of $\ln (t)$ with four sets of parameters ($\protect\mu $ and $%
\Delta $) indicated by dots a-d as shown in Fig. 1(a) . $\protect\delta 
\protect\eta (t)$ is defined in Eq. (\protect\ref{dnt}). The slope of the
green line in (a) and (b) is 0.5 while in (c) and (d) is 1.}
\label{fig}
\end{figure*}

\begin{figure}[t]
\centering \includegraphics[width=0.49\textwidth]{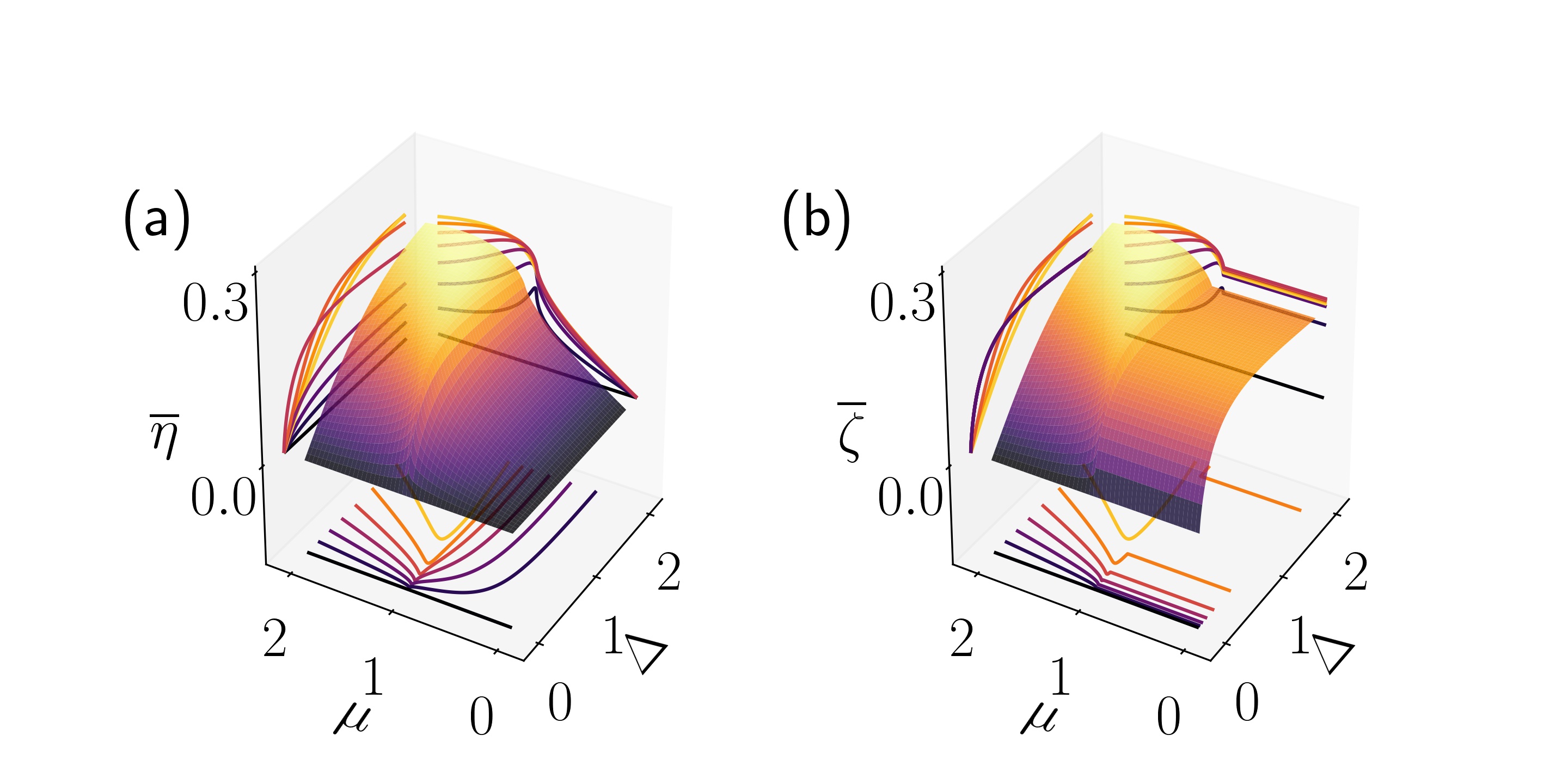} 
\caption{Three-dimensional plots of the average order parameters (a) $\overline{\protect\eta }$ and (b) $\overline{\protect\zeta }$ obtained from
Eqs. (\protect\ref{eta_sum}) and (\protect\ref{zeta_sum})\ as the function
of $\protect\mu $ and $\Delta $ for case $N=4000$ in the first quadrant. The
patterns are not dependent on the quadrant. The two quantities vary, but
both can identify the phase boundary at $\protect\mu =1$\ and $\Delta =0$ $(\left\vert \protect\mu \right\vert <1)$ (also refer to Fig. \protect\ref{fig5}). The other parameter is $J=1$.}
\label{fig4}
\end{figure}

\begin{figure*}[tbh]
\centering \includegraphics[width=1\textwidth]{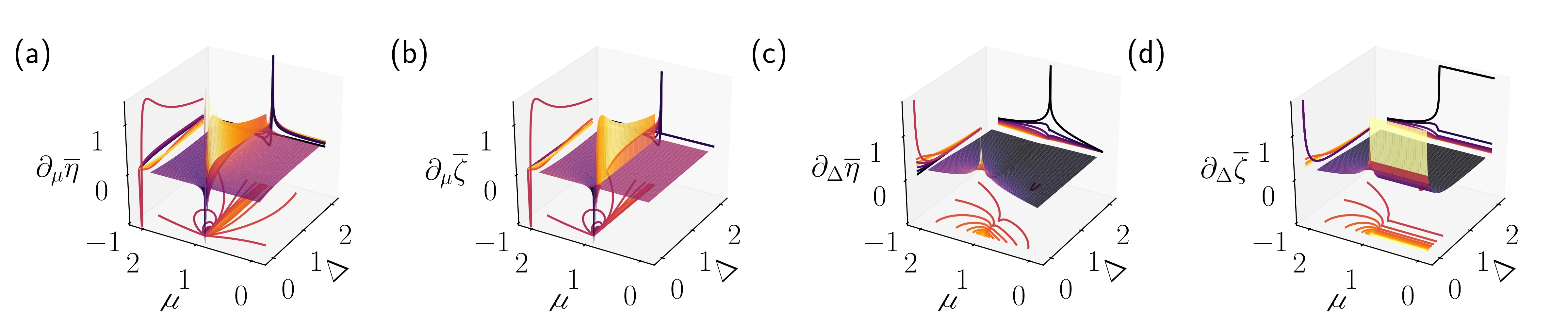}
\caption{Three-dimensional plots of (a) $\partial _{\protect\mu }\overline{%
\protect\eta }$, (b) $\partial _{\protect\mu }\overline{\protect\zeta }$,\
(c) $\partial _{\Delta }\overline{\protect\eta }$ and (d) $\partial _{\Delta
}\overline{\protect\zeta }$ as the function of $\protect\mu $ and $\Delta $
for case $N=4000$ in the first quadrant; the forms of $\overline{\protect%
\eta }$ and $\overline{\protect\zeta }$ are obtained from Eqs. (\protect\ref%
{eta_sum}) and (\protect\ref{zeta_sum}). The aforementioned four quantities
can identify the phase boundary at $\protect\mu =1$\ and $\Delta =0$ $%
(\left\vert \protect\mu \right\vert <1)$. The other parameter is $J=1$. }
\label{fig5}
\end{figure*}

\begin{figure*}[tbh]
\centering \includegraphics[width=1\textwidth]{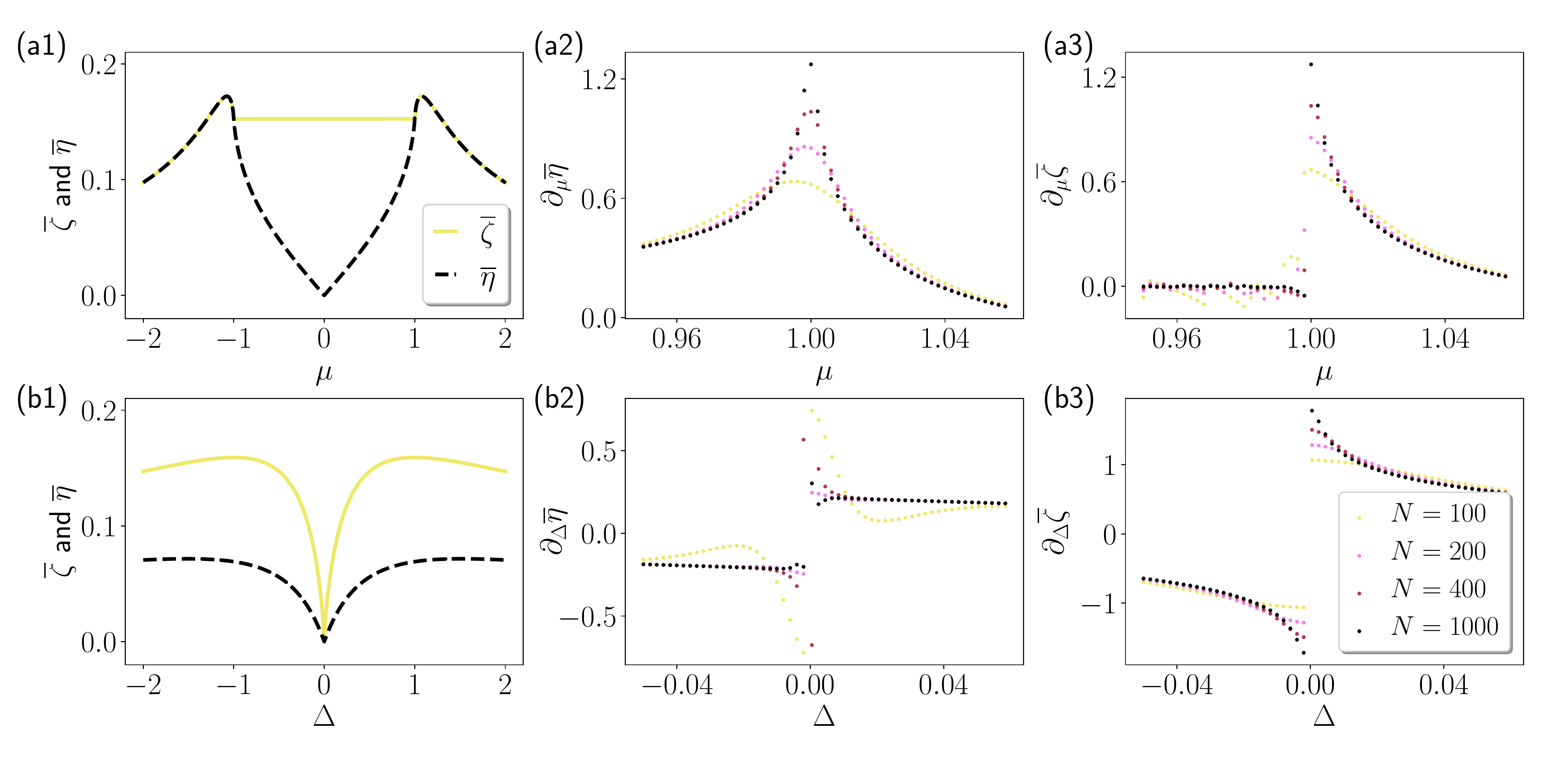}
\caption{Profiles of $\overline{\protect\zeta }$ and $\overline{\protect\eta 
}$ obtained from Eqs. (\protect\ref{eta_sum}) and (\protect\ref{zeta_sum})
at lines (a1) $\Delta =0.6$ and (b1) $\protect\mu =0.6$. The corresponding
partial derivatives of $\overline{\protect\eta }$ and $\overline{\protect%
\zeta }$ (a2) $\partial _{\protect\mu }\overline{\protect\eta }$, (a3) $%
\partial _{\protect\mu }\overline{\protect\zeta }$, (b2) $\partial _{\Delta }%
\overline{\protect\eta }$ and (b3) $\partial _{\Delta }\overline{\protect%
\zeta }$ for finite sizes $N=100$, $200$, $400$ and $1000,$ respectively. In
(a2) and (a3), the values of $\partial _{\protect\mu }\overline{\protect\eta 
}\left( \protect\mu =1\right) $ and $\partial _{\protect\mu }\overline{%
\protect\zeta }\left( \protect\mu =1^{+}\right) $ increase rapidly with
increasing $N$, indicating the divergence of $\partial _{\protect\mu }%
\overline{\protect\eta }$ and $\partial _{\protect\mu }\overline{\protect%
\zeta }$ at phase boundary $\left\vert \protect\mu \right\vert =1$ in the
large $N$ limit. (b2) and (b3) indicate that $\partial _{\Delta }\overline{%
\protect\eta }$ is discontinuous and $\partial _{\Delta }\overline{\protect%
\zeta }$ is divergent at the phase boundary $\left\vert \Delta \right\vert
=0 $ in the large $N$ limit. The other parameter used for numerical
calculations is $J=1$.}
\label{fig6}
\end{figure*}

\section{Model and order parameters}

\label{Model and order parameters}

We consider the following fermionic Hamiltonian on a lattice of length $N$

\begin{eqnarray}
H &=&\sum\limits_{j=1}^{N}[-Jc_{j}^{\dag }c_{j+1}-\Delta c_{j}^{\dag
}c_{j+1}^{\dag }+\mathrm{H.c.}  \notag \\
&&+\mu \left( 2n_{j}-1\right) ],  \label{H}
\end{eqnarray}%
where $c_{j}^{\dag }$ $(c_{j})$\ is a fermionic creation (annihilation)
operator on site $j$, $n_{j}=c_{j}^{\dag }c_{j}$, $J$ the tunneling rate, $%
\mu
$ the chemical potential, and $\Delta $\ the strength of the $p$-wave pair
creation (annihilation). $c_{N+1}=c_{1}$ is defined for periodic boundary
condition. The Hamiltonian in Eq. (\ref{H}) has a rich phase diagram that
describes a spin-polarized $p$-wave superconductor in one dimension. This
system has a topological phase in which a zero- energy Majorana mode is
located at each end of a long chain. It is the fermionized version of the
well-known one-dimensional transverse-field Ising model \cite{Pfeuty}, which
is one of the simplest solvable models that exhibits quantum criticality and
phase transition with spontaneous symmetry breaking \cite{SachdevBook}%
.Several studies have been conducted with a focus on long-range Kitaev
chains, in which the superconducting pairing term decays with distance as a
power law \cite{TPCHOY,AGG,DV1,DV2,OV,LL,UB}.

This study centered on the dynamics of a system with NN superconducting
pairing term, which warrants further-systematic studies, such as those on
long-range correlations. The Hamiltonian can be exactly solved
based on Fourier transformation (see Appendix \ref{AppendixA}). The
ground-state wave function can be expressed as%
\begin{equation}
\left\vert \text{\textrm{G}}\right\rangle =\prod_{\pi >k>0}\left\vert
\varphi _{k\mathrm{e}}^{-}\right\rangle ,
\end{equation}%
with groundstate energy 
\begin{equation}
E_{\mathrm{g}}=-\sum_{\pi >k>0}\varepsilon _{k}.
\end{equation}%
Accordingly, the phase diagram can be obtained from $\left\vert \text{%
\textrm{G}}\right\rangle $\ and $E_{\mathrm{g}}$, with the phase boundaries 
\begin{equation}
\mu =\pm 1\text{, and }\Delta =0\text{, for }\left\vert \mu \right\vert <1.
\end{equation}%
These lines separate four regions in the parameter space, representing
different phases with winding numbers \cite{CKC,NL} $N=0$, and $\pm 1$,
respectively. Fig. \ref{fig1}(a) depicts the phase diagram. The phase
diagram indicates that the chemical potential $\mu $ has a subtle role in
the ground state. It restricts the value of $\left\vert \mu \right\vert $\
to maintain the non-trivial phases. The might be because the sufficiently
large $\left\vert \mu \right\vert $\ suppresses pair creation. Thus the
number of pairs is higher in nontrivial phases than in trivial phases which
leads to the following two questions. First, how to quantify the amount of
pairs in each phase? Second, what are the features of the pairing in real or
momentum space. The answers are the main part of our goals in this work.

To answer the questions, we explore the features of a
superconducting state by introducing two types of order parameters \cite{BP}%
. Two operators $\widehat{\eta }_{l}$\ and $\widehat{\zeta }_{k}$\
characterize pairing channels in real space with%
\begin{equation}
\widehat{\eta }_{l}=\frac{2}{\pi }\sum\limits_{\text{odd }r>0}\frac{1}{r}%
\left( c_{l}c_{l+r}-c_{l}^{\dag }c_{l+r}^{\dag }\right) ,
\end{equation}%
and $k$\ space with%
\begin{equation}
\widehat{\zeta }_{k}=i\left( c_{-k}c_{k}-c_{k}^{\dag }c_{-k}^{\dag }\right) .
\end{equation}%
For a given state $\left\vert \psi \right\rangle $, the quantity $%
|\left\langle \psi \right\vert \widehat{\eta }_{l}\left\vert \psi
\right\rangle |$\ helps measure the rate of transition for a pair located
around the $l$th site, whereas $|\left\langle \psi \right\vert \widehat{%
\zeta }_{k}\left\vert \psi \right\rangle |$\ helps measure the rate of
transition for a pair at the $k$\ channel. The corresponding order
parameters are defined on the basis of the average magnitude over all
channels, as expressed in%
\begin{equation}
\eta =\frac{1}{N}\sum_{l}|\left\langle \psi \right\vert \widehat{\eta }%
_{l}\left\vert \psi \right\rangle |,
\end{equation}%
and%
\begin{equation}
\zeta =\frac{1}{N}\sum_{\pi >k>0}|\left\langle \psi \right\vert \widehat{%
\zeta }_{k}\left\vert \psi \right\rangle |.
\end{equation}%
Nonzero $\eta $\ and $\zeta $\ indicate that the state $\left\vert \psi
\right\rangle $\ is a superconducting state. In general, $\eta $\ and $\zeta 
$\ have different values for a given state. In the following subsections, we
describe the analytical expressions of $\eta $\ and $\zeta $\ for the ground
states of the system in different regions and their behaviors at phase
boundaries.

\subsection{Pairing in $k$ space}

\label{K space} The order parameter of the ground state is expressed as%
\begin{equation}
\zeta _{\mathrm{g}}=\frac{1}{N}\sum_{\pi >k>0}\left\vert \left\langle \text{%
\textrm{G}}\right\vert \widehat{\zeta }_{k}\left\vert \text{\textrm{G}}%
\right\rangle \right\vert =\frac{1}{N}\sum_{\pi >k>0}\left\vert \left\langle
\varphi _{k\mathrm{e}}^{-}\right\vert \widehat{\zeta }_{k}\left\vert \varphi
_{k\mathrm{e}}^{-}\right\rangle \right\vert .
\end{equation}%
Direct derivation shows that the order parameter is%
\begin{equation}
\zeta _{\mathrm{g}}=\frac{2\left\vert \Delta \right\vert }{N}\sum_{\pi
>k>0}\left\vert \frac{\sin k}{\varepsilon _{k}}\right\vert =\frac{\left\vert
\Delta \right\vert }{\pi }\int_{0}^{\pi }\frac{\sin k}{\varepsilon _{k}}%
\mathrm{d}k  \label{zeta_g}
\end{equation}%
in the large $N$ limit, which can be expressed explicitly in the following
regions. (i) When $\left\vert \mu \right\vert >1$, the order parameter is%
\begin{equation}
\zeta _{\mathrm{g}}=\frac{\left\vert \Delta \right\vert }{2\pi \sqrt{%
1-\Delta ^{2}}}\ln \left\vert \frac{\left\vert \mu \right\vert +\sqrt{%
1-\Delta ^{2}}}{\left\vert \mu \right\vert -\sqrt{1-\Delta ^{2}}}\right\vert
\end{equation}%
for $\left\vert \Delta \right\vert \neq 1$ but $\left( \left\vert \mu
\right\vert \pi \right) ^{-1}$\ at $\left\vert \Delta \right\vert =1$.\ (ii)
When $\left\vert \mu \right\vert \leqslant 1$, the order parameter is%
\begin{equation}
\zeta _{\mathrm{g}}=\frac{\left\vert \Delta \right\vert }{2\pi \sqrt{%
1-\Delta ^{2}}}\ln \left\vert \frac{1+\sqrt{1-\Delta ^{2}}}{1-\sqrt{1-\Delta
^{2}}}\right\vert  \label{mu<1}
\end{equation}%
for $\left\vert \Delta \right\vert \neq 1$ but $\pi ^{-1}$\ at $\left\vert
\Delta \right\vert =1$. This suggests that when $\left\vert \mu \right\vert
\leqslant 1,$ $\zeta _{\mathrm{g}}$ is not dependent on $\mu $.
Unexpectedly, the number of pairs is determined using only the parameter $%
\Delta $\ in non-trivial phases. The underlying mechanism is an open
question in this work. By contrast, when $\left\vert \mu \right\vert >1$,\ $%
\zeta _{\mathrm{g}}$\ decays as $\left\vert \mu \right\vert $\ increases.
Specifically, it decays linearly with $\left\vert \mu \right\vert -1$ for
small $\left\vert \mu \right\vert -1$ and as $\left\vert \mu \right\vert
^{-1}$\ for large $\left\vert \mu \right\vert $.

Nevertheless, we can only understand this result mathematically
from an example in classical Electromagnetics. We consider a charged conducting
ellipsoid with the surface equations%
\begin{equation}
(x-\mu )^{2}+\frac{y^{2}+z^{2}}{\Delta ^{2}}=1.
\end{equation}%
The total charge is $Q=4\left\vert \Delta \right\vert $. Then surface charge
density distribution along the $x$\ axis is%
\begin{equation}
\sigma =\frac{1}{\pi \left\vert \Delta \right\vert }\left[ (x-\mu )^{2}+%
\frac{R^{2}}{\Delta ^{4}}\right] ^{-\frac{1}{2}}.
\end{equation}%
And the electric potential at the origin is%
\begin{equation}
\Phi =\frac{\left\vert \Delta \right\vert }{2\pi }\int_{0}^{\pi }\frac{\sin
\theta }{\sqrt{\left( \Delta \sin \theta \right) ^{2}+(\cos \theta -\mu )^{2}%
}}d\theta ,
\end{equation}%
where 
\begin{equation}
\tan \theta =\frac{R}{\Delta \left( x-\mu \right) }.
\end{equation}%
\ Regardless of the integration, the physics tells us that $\Phi $\ is
always constant when the origin lies inside the ellipsoid ($\left\vert \mu
\right\vert <1$). We note that the expression of $\Phi $\ is identical to
the Eq. (\ref{zeta_g}).

\subsection{Pairing in real space}

\label{R space} The following order parameter%
\begin{equation}
\eta _{\mathrm{g}}=\frac{1}{N}\sum_{l}\left\vert \left\langle \text{\textrm{G%
}}\right\vert \widehat{\eta }_{l}\left\vert \text{\textrm{G}}\right\rangle
\right\vert
\end{equation}%
is identical to $\zeta _{\mathrm{g}}$. The fact that $\left\langle \text{%
\textrm{G}}\right\vert \widehat{\eta }_{l}\left\vert \text{\textrm{G}}%
\right\rangle =\left\langle \text{\textrm{G}}\right\vert \widehat{\eta }%
_{l+1}\left\vert \text{\textrm{G}}\right\rangle $ indicates that%
\begin{equation}
\eta _{\mathrm{g}}=\left\vert \left\langle \text{\textrm{G}}\right\vert
\sum_{l}\frac{1}{N}\widehat{\eta }_{l}\left\vert \text{\textrm{G}}%
\right\rangle \right\vert .
\end{equation}%
Applying Fourier transformation in Eq. (\ref{FT}) in the large $N$ limit,
the following equation can be obtained: 
\begin{equation}
\eta _{\mathrm{g}}=\zeta _{\mathrm{g}}=\frac{1}{N}\sum_{\pi >k>0}\left\vert
\left\langle \text{\textrm{G}}\right\vert i\left( c_{-k}c_{k}-c_{k}^{\dag
}c_{-k}^{\dag }\right) \left\vert \text{\textrm{G}}\right\rangle \right\vert
.
\end{equation}%
This is the primary finding of the present study. It appears that the ground
state favors a balance between the two types of pairing. However, not all
states can maintain such a balance. Thus, the aformentioned queries are
resolved. Regarding the first query, two types of order parameters are
deliberately selected to quantify the number of pairs in two aspects.
Regarding the second query, the equality of the two quantities $\eta _{%
\mathrm{g}}$ and $\zeta _{\mathrm{g}}$ for the ground state itself is a key
feature of the ground state.

In particular, the analytical behavior of $\eta _{\mathrm{g}}$\
($\zeta _{\mathrm{g}}$) as function of $\left( \Delta ,\mu \right) $\ can
identify the phase diagram (see Appendix \ref{AppendixB}). It is found that
second-order quantum phase transitions occur at phase boundaries. The
first-order derivatives of two parameters with respect to $\Delta $\ or $%
\mu $\ are discontinuous, and then result in the divergence of the
corresponding second-order derivatives. The profiles of $\zeta
_{\mathrm{g}}\left( \mu ,\Delta \right) $\ at the line $\Delta =0.6$\ and $%
\mu =0.6$\ are presented in Figs. \ref{fig1}(b) and \ref{fig1}(c),
respectively. $\zeta _{\mathrm{g}}$\ and its partial derivatives with
respect to $\mu $\ and $\Delta $\ for a finite $N$\ lattice in the first
quadrant of the $\mu -\Delta $\ plane are shown in Figs. \ref{fig2}(a)-\ref%
{fig2}(c). Numerical data for the finite system reveal that the divergence
at phase boundaries is demonstrated by peaks. 

\section{Non-equilibrium \textbf{stationary} state}

\label{Dynamics}

In this section, we present the features of the dynamic behavior of the
system. It relates to the time evolution $\left\vert \psi (t)\right\rangle $
for a particular initial state

\begin{equation}
\left\vert \psi (0)\right\rangle =\prod_{\pi >k>0}\left\vert 0\right\rangle
_{k}\left\vert 0\right\rangle _{-k},
\end{equation}%
which is essentially an empty state $\prod_{j=1}^{N}\left\vert
0\right\rangle _{j}$ for fermion in real space. The evolved state%
\begin{eqnarray}
\left\vert \psi (t)\right\rangle &=&e^{-iHt}\left\vert \psi (0)\right\rangle
=\prod_{\pi >k>0}e^{-iH_{k}t}\left\vert 0\right\rangle _{k}\left\vert
0\right\rangle _{-k}  \notag \\
&=&\prod_{\pi >k>0}\frac{\sin \left( \varepsilon _{k}t\right) }{\varepsilon
_{k}}\{-2\Delta \sin k\left\vert 1\right\rangle _{k}\left\vert
1\right\rangle _{-k}  \notag \\
&&+\left[ \varepsilon _{k}\cot \left( \varepsilon _{k}t\right) -i2\left(
\cos k-\mu \right) \right] \left\vert 0\right\rangle _{k}\left\vert
0\right\rangle _{-k}\},
\end{eqnarray}%
may trend toward a stationary state after a sufficiently long
time. The non-equilibrium stationary state driven by $H$ is
intimately associated with the phase diagram of the ground state (Fig. \ref%
{fig1}(a)). We focus on the time-dependent order parameters for the evolved
state $\left\vert \psi (t)\right\rangle $. Based on the fact that $%
\left\vert \psi (t)\right\rangle $\ represents the eigenstates of the
operator $T$, the following explicit forms can be obtained: 
\begin{eqnarray}
\eta (t) &=&\frac{1}{N}\sum_{l}\left\vert \left\langle \psi (t)\right\vert 
\widehat{\eta }_{l}\left\vert \psi (t)\right\rangle \right\vert  \notag \\
&=&\frac{2}{N}\left\vert \sum_{\pi >k>0}F(k)\sin ^{2}\left( \varepsilon
_{k}t\right) \right\vert ,  \label{etaT}
\end{eqnarray}%
and%
\begin{eqnarray}
\zeta (t) &=&\frac{1}{N}\sum_{\pi >k>0}\left\vert \left\langle \psi
(t)\right\vert \widehat{\zeta }_{k}\left\vert \psi (t)\right\rangle
\right\vert  \notag \\
&=&\frac{2}{N}\sum_{\pi >k>0}\left\vert F(k)\right\vert \sin ^{2}\left(
\varepsilon _{k}t\right) ,  \label{zetaT}
\end{eqnarray}%
where the key function is%
\begin{equation}
F(k)=\frac{\left( \cos k-\mu \right) \Delta \sin k}{\left( \cos k-\mu
\right) ^{2}+\Delta ^{2}\sin ^{2}k}.
\end{equation}

Figs. \ref{fig3}(b1) and \ref{fig3}(b2) present $\zeta (t)$\
and $\eta (t)$\ as the function of time for finite-size systems with several
typical system parameters $\left( \mu ,\Delta \right) $. As expected, both
quantities become steady after a long period. $\zeta (t)$\ and $\eta (t)$\
behave as damping oscillations around a certain equilibrium position. The
quasi frequencies and decay rates can be evaluated through a saddle point
integration in the continuum limit \cite{AS,AAM,SA}. We rewrite $\eta \left(
t\right) $\ in the form%
\begin{equation}
\eta \left( t\right) =\left\vert \eta \left( \infty \right) -\delta \eta
\left( t\right) \right\vert
\end{equation}%
where 
\begin{eqnarray}
\eta \left( \infty \right) &=&\frac{1}{2\pi }\int_{0}^{\pi }F\left( k\right) 
\mathrm{d}k, \\
\delta \eta \left( t\right) &=&\frac{1}{2\pi }\int_{0}^{\pi }F\left(
k\right) \cos \left( 2\varepsilon _{k}t\right) \mathrm{d}k.  \label{dnt}
\end{eqnarray}%
The results depends on the location $k_{0}$\ of the extrema $\varepsilon
_{k} $. (i) In this case with $k_{0}\neq 0,\pi $, we have quasi frequency $%
\omega \approx 2\varepsilon _{k_{0}}$\ and $\delta \eta \left( t\right) 
 \sim t^{-0.5}$. (ii) In this case with $k_{0}=0,\pi $, we have quasi
frequency $\omega \approx 2\varepsilon _{k_{0}=0}$\ and $\delta \eta \left(
t\right) \sim t^{-1}$. Similar analysis can be applied to $\zeta
\left( t\right) $. These results accord with the plots in Fig. \ref{fig}. 

On the other hand, such behaviors can also be understood by a simple
picture, when the spectrum contains higher degeneracy, which relates to the
peak of the density of state (DOS) of the system. The DOS at energy level $E$%
\ is generally defined as%
\begin{equation}
D\left( E\right) =\lambda \frac{\text{d}N\left( E\right) }{\text{d}E}
\end{equation}%
where d$N\left( E\right) $\ indicates the number of energy levels that
appear in the interval $\left[ E,E+dE\right] ,$\ with the normalization
factor $\lambda =\frac{1}{N}$. Figs. \ref{fig3}(c1)-\ref{fig3}(c4), depict
the energy levels and DOS as the functions of momentum $k$\ and $E$,
respectively. When the peaks of $D\left( E\right) $\ appear, the quasi
frequency is related to the energy band gap $\delta E$, which is the
distance between two peaks.

No matter what mechanism of the decay, $\zeta (t)$\ and $\eta (t)$\ approach
to their stable values, which can be obtained from%
\begin{equation}
\overline{\eta }=\lim_{T\rightarrow \infty }\frac{1}{T}\int_{0}^{T}\eta (t)%
\mathrm{d}t=\frac{1}{N}\left\vert \sum_{\pi >k>0}F(k)\right\vert ,
\label{eta_sum}
\end{equation}%
and%
\begin{equation}
\overline{\zeta }=\lim_{T\rightarrow \infty }\frac{1}{T}\int_{0}^{T}\zeta (t)%
\mathrm{d}t=\frac{1}{N}\sum_{\pi >k>0}\left\vert F(k)\right\vert .
\label{zeta_sum}
\end{equation}%
It is two types of summation for the same function $F(k)$. In the large $N$\
limit, $\overline{\eta }$\ and $\overline{\zeta }$\ can be expressed as the
integral equations:%
\begin{equation}
\overline{\eta }=\frac{1}{2\pi }\left\vert \int_{0}^{\pi }F(k)\mathrm{d}%
k\right\vert ,
\end{equation}%
and

\begin{equation}
\overline{\zeta }=\frac{1}{2\pi }\int_{0}^{\pi }\left\vert F(k)\right\vert 
\mathrm{d}k\mathbf{.}
\end{equation}%
Straightforward derivations lead to the development of analytical expressions

\begin{equation}
\overline{\eta }=\frac{\left\vert \Delta \right\vert }{2\pi \left( 1-\Delta
^{2}\right) }\ln \left\vert \frac{\left\vert \mu \right\vert +1}{\left\vert
\mu \right\vert -1}\right\vert -\Lambda \left( \mu ,\Delta \right) ,
\label{eta}
\end{equation}%
where $\Lambda =\mu \Delta \pi ^{-1}\left( 1-\Delta ^{2}\right) ^{-1}$ at $%
\mu ^{2}+\Delta ^{2}=1$, and%
\begin{eqnarray}
\Lambda \left( \mu ,\Delta \right) &=&\frac{\Delta ^{2}\left\vert \mu
\right\vert }{2\pi \left( 1-\Delta ^{2}\right) \sqrt{\mu ^{2}+\Delta ^{2}-1}}
\notag \\
&&\times \ln \left\vert \frac{\sqrt{\mu ^{2}+\Delta ^{2}-1}+\left\vert
\Delta \right\vert }{\sqrt{\mu ^{2}+\Delta ^{2}-1}-\left\vert \Delta
\right\vert }\right\vert ,
\end{eqnarray}%
when $\mu ^{2}+\Delta ^{2}\neq 1$. Conversely,

\begin{equation}
\overline{\zeta }=\left\{ 
\begin{array}{cc}
\overline{\eta } & \left\vert \mu \right\vert \geqslant 1 \\ 
\frac{\left\vert \Delta \right\vert \ln \left( \left\vert \Delta \right\vert
\right) }{\pi \left( \Delta ^{2}-1\right) } & \left\vert \mu \right\vert <1%
\end{array}%
\right. .  \label{zeta}
\end{equation}
 the time-averaged analytical expressions for $\eta$ and $\zeta$ obtained from Eqs. (\ref{eta}) and (\ref{zeta}) are shown graphically in Fig. \ref{fig3}(a).

 Similar to the static order parameters, analytical behavior of $%
\overline{\zeta }$\ and $\overline{\eta }$\ as function of $\left( \Delta
,\mu \right) $ can still identify the phase diagram (see Appendix \ref%
{AppendixB}). Fig. \ref{fig4} shows the average order parameters $\overline{%
\zeta }$\ and $\overline{\eta }$\ obtained from Eqs. (\ref{eta_sum}) and (%
\ref{zeta_sum}) as the function of $\mu $\ and $\Delta $\ for the case $%
N=4000$\ in the first quadrant of the $\mu -\Delta $\ plane. Fig. \ref{fig5}
presents $\partial _{\mu }\overline{\eta },$\ $\partial _{\mu }\overline{%
\zeta }$, $\partial _{\Delta }\overline{\eta },$\ and $\partial _{\Delta }%
\overline{\zeta }$, the partial dervatives of $\overline{\zeta }$\ and $%
\overline{\eta }$\ with respect to $\mu $\ and $\Delta $, for finite system.
Fig. \ref{fig6}(a1)-\ref{fig6}(a3) show the profiles of $%
\overline{\zeta }\left( \mu ,\Delta \right) $, $\overline{\eta }\left( \mu
,\Delta \right) $, $\partial _{\mu }\overline{\eta }$\ and $\partial _{\mu }%
\overline{\zeta }$\ at the line $\Delta =0.6.$\ Fig. \ref{fig6}(b1)-\ref%
{fig6}(b3) present the profiles of $\overline{\zeta }\left( \mu ,\Delta
\right) $, $\overline{\eta }\left( \mu ,\Delta \right) $, $\partial _{\Delta
}\overline{\eta }$\ and $\partial _{\Delta }\overline{\zeta }$\ at the line $%
\mu =0.6$, indicating that four\ quantities, $\partial _{\mu }\overline{\eta 
},$\ $\partial _{\mu }\overline{\zeta }$, $\partial _{\Delta }\overline{\eta 
},$\ and $\partial _{\Delta }\overline{\zeta }$, can identify the phase
boundary at $\mu =1$\ and $\Delta =0$\ $(\mu <1)$. Compared with $\eta _{%
\mathrm{g}}$\ and $\zeta _{\mathrm{g}},$\ which are always identical in all
regions, $\overline{\zeta }$\ and $\overline{\eta }$\ vary in non-trivial
regions, providing quantitative description of the pairing mechanism for a
non-equilibrium state. Moreover, the non-equilibrium state is also a
superconducting state because the values of $\overline{\zeta }$\ and $%
\overline{\eta }$\ have the same order as those of $\zeta _{\mathrm{g}}$\
and $\eta _{\mathrm{g}}$. The fact that $\overline{\zeta }>\overline{\eta }$%
\ indicates that BCS-like pairing in momentum space is favored. Before
ending this work, we would like to point out that the obtained results
depend on the initial state. For example, consider an initial state in the
form%
\begin{equation}
\left\vert \psi (0)\right\rangle =\prod\limits_{k\in \left\{ k_{a}\right\}
}\left\vert 1\right\rangle _{k}\left\vert 0\right\rangle
_{-k}\prod\limits_{k\in \left\{ k_{b}\right\} }\left\vert 0\right\rangle
_{k}\left\vert 1\right\rangle _{-k},
\end{equation}%
with $\left\{ k_{a}\right\} \oplus \left\{ k_{b}\right\} \in \left( 0,\pi
\right) $,\ which is the eigenstate of $H$\ with arbitrary parameters $%
\left( J,\Delta ,\mu \right) $\ due to the fact%
\begin{equation}
H_{k}\left\vert 1\right\rangle _{k}\left\vert 0\right\rangle
_{-k}=H_{k}\left\vert 0\right\rangle _{k}\left\vert 1\right\rangle _{-k}=0,
\end{equation}%
from the Appendix \ref{AppendixA}. We find that%
\begin{equation}
\sum_{l}\widehat{\eta }_{l}\left\vert \psi (0)\right\rangle =\sum_{\pi >k>0}%
\widehat{\zeta }_{k}\left\vert \psi (0)\right\rangle =0,
\end{equation}%
which result in $\overline{\eta }=\overline{\zeta }=0$, indicating that
quantities $\overline{\eta }$\ and $\overline{\zeta }\ $cannot indicate the
phase diagram.

\section{Summary}

\label{sec_summary}

In summary, we have studied the quench dynamics of a Kitaev chain by
introducing two types of order parameters that characterize two different
pairing mechanisms, namely local pairing in real space and BCS-like pairing
in momentum space. We have shown exactly that the ground states over the
entire parameter region support the identical values of them, indicating a
balance between the two types of pairing channels; This is an unprecedented
feature of the ground state. The balance is disturbed in a non-equilibrium
state when the post-quench Hamiltonian is in a topologically non-trivial
phase, whereas it is maintained in a topologically trivial phase.
Furthermore, non-equilibrium superconducting states favor the BCS-like
pairing. ---Our findings present an alternative approach to generate a
superconducting state from a trivial empty state and illuminate the pairing
mechanism.

\section*{Acknowledgment}

We acknowledge the support of NSFC (Grants No. 11874225).

\appendix
\section{Solution and phase diagram}

\label{AppendixA} \setcounter{equation}{0} \renewcommand{\theequation}{A%
\arabic{equation}} \renewcommand{\thesubsection}{\arabic{subsection}}

The Hamiltonian (\ref{H}) is exactly solvable because of the translational
symmetry of the system, i.e., $\left[ T,H\right] =0$,\ where $T$\ is the
translational operator, which is defined by $Tc_{l}T^{-1}=c_{l+1}$. Taking
the Fourier transformation

\begin{equation}
c_{j}=\frac{1}{\sqrt{N}}\sum\limits_{k}e^{ikj}c_{k},  \label{FT}
\end{equation}%
of the Hamiltonian (\ref{H}), with wave vector $k\in (-\pi ,\pi ]$, we can
obtain the following equation: 
\begin{eqnarray}
H &=&\sum_{k}[\left( -2J\cos k+2\mu \right) c_{k}^{\dag }c_{k}  \notag \\
&&+i\Delta \sin k\left( c_{-k}c_{k}+c_{-k}^{\dag }c_{k}^{\dag }\right) -\mu
].
\end{eqnarray}%
For ease of further analysis, the $J$\ value can be considered to be $1$ and
the Hamiltonian can be expressed through the Nambu representation 
\begin{eqnarray}
H &=&\sum_{\pi >k>0}H_{k}, \\
H_{k} &=&-2\left( 
\begin{array}{cc}
c_{k}^{\dag } & c_{-k}%
\end{array}%
\right) \left( 
\begin{array}{cc}
\cos k-\mu & i\Delta \sin k \\ 
-i\Delta \sin k & \mu -\cos k%
\end{array}%
\right) \left( 
\begin{array}{c}
c_{k} \\ 
c_{-k}^{\dag }%
\end{array}%
\right) ,
\end{eqnarray}%
where the Hamiltonian $H_{k}$\ in each invariant subspace satisfies the
commutation relation%
\begin{equation}
\left[ H_{k},H_{k^{\prime }}\right] =0.
\end{equation}%
This allows us to treat the diagonalization and dynamics governed by $H_{k}$%
\ individually. For a given $k$, the Hamiltonian $H_{k}$\ in the basis ($%
\left\vert 0\right\rangle _{k}\left\vert 0\right\rangle _{-k},\left\vert
1\right\rangle _{k}\left\vert 1\right\rangle _{-k},\left\vert 1\right\rangle
_{k}\left\vert 0\right\rangle _{-k},\left\vert 0\right\rangle _{k}\left\vert
1\right\rangle _{-k}$) is expressed as a $4\times 4$ matrix%
\begin{equation}
h_{k}=2\left( 
\begin{array}{cccc}
\cos k-\mu & i\Delta \sin k & 0 & 0 \\ 
-i\Delta \sin k & \mu -\cos k & 0 & 0 \\ 
0 & 0 & 0 & 0 \\ 
0 & 0 & 0 & 0%
\end{array}%
\right) ,
\end{equation}%
where $\left\vert 1\right\rangle _{k}=c_{k}^{\dag }\left\vert 0\right\rangle
_{k}$ and $c_{k}\left\vert 0\right\rangle _{k}=0$. The eigenstates $%
\left\vert \varphi _{k\mathrm{\lambda }}^{\pm }\right\rangle $\ [$\mathrm{%
\lambda }=\mathrm{e}$ $\mathrm{(o)}$ denotes the even (odd) parity of the
particle number] are

\begin{eqnarray}
\left\vert \varphi _{k\mathrm{e}}^{\pm }\right\rangle &=&\frac{1}{\Omega
^{\pm }}\left( a_{k}^{\pm }\left\vert 0\right\rangle _{k}\left\vert
0\right\rangle _{-k}+\left\vert 1\right\rangle _{k}\left\vert 1\right\rangle
_{-k}\right) , \\
\left\vert \varphi _{k\mathrm{o}}^{+}\right\rangle &=&\left\vert
1\right\rangle _{k}\left\vert 0\right\rangle _{-k},\left\vert \varphi _{k%
\mathrm{o}}^{-}\right\rangle =\left\vert 0\right\rangle _{k}\left\vert
1\right\rangle _{-k},
\end{eqnarray}%
where $\Omega ^{\pm }=\sqrt{1+\left\vert a_{k}^{\pm }\right\vert ^{2}}$ is
the normalization coefficient in the context of Dirac inner product with%
\begin{equation}
a_{k}^{\pm }=i\frac{\cos k-\mu +\epsilon _{k\mathrm{e}}^{\pm }/2}{\Delta
\sin k},
\end{equation}%
and corresponding energies are

\begin{equation}
\epsilon _{k\mathrm{e}}^{\pm }=\pm \varepsilon _{k},\epsilon _{k\mathrm{o}%
}^{\pm }=0,
\end{equation}%
with%
\begin{equation}
\varepsilon _{k}=2\sqrt{\left( \mu -\cos k\right) ^{2}+\Delta ^{2}\sin ^{2}k}%
.
\end{equation}%
Accordingly, the ground-state wave function can be expressed as%
\begin{equation}
\left\vert \text{\textrm{G}}\right\rangle =\prod_{\pi >k>0}\left\vert
\varphi _{k\mathrm{e}}^{-}\right\rangle .
\end{equation}%
$\left\vert \text{\textrm{G}}\right\rangle $\ is the eigenstate of $T$,
obeying $T\left\vert \text{\textrm{G}}\right\rangle =e^{i\Omega }\left\vert 
\text{\textrm{G}}\right\rangle $, where $\Omega $\ is a real number.\
Furthermore, the phase-transition lines (phase boundary) can be determined
using min$(\varepsilon _{k})=0$, which results in the gapless lines 
\begin{equation}
\mu =\pm 1\text{, and }\Delta =0\text{, for }\left\vert \mu \right\vert <1.
\end{equation}%
These lines separate four regions in the parameter space, representing
different phases with winding numbers \cite{CKC,NL} $N=0$, and $\pm 1$,
respectively.

\section{Analytical behavior of static and dynamic order parameters}

\label{AppendixB} \setcounter{equation}{0} \renewcommand{\theequation}{B	%
\arabic{equation}} \renewcommand{\thesubsection}{\arabic{subsection}} The
analytical behavior of $\eta _{\mathrm{g}}$ ($\zeta _{\mathrm{g}}$) as the
function of $\left( \Delta \text{ and }\mu \right) $\ can identify the phase
diagram. On the basis of the aforementioned euqations, we found that
second-order quantum phase transitions occur at phase boundaries. Around
lines $\left\vert \mu \right\vert =1$, we have $\partial _{\mu }\zeta _{%
\mathrm{g}}=0$ at $\mu =\pm 1^{\mp }$ but $\partial _{\mu }\zeta _{\mathrm{g}%
}=\mp 2\pi ^{-1}\Delta ^{-1}$ at $\mu =\pm 1^{\pm }$. Then, $\partial _{\mu
}^{2}\zeta _{\mathrm{g}}$ diverges at these two lines. We have 
\begin{eqnarray}
\zeta _{\mathrm{g}}|_{\Delta \rightarrow 0} &\approx &-2\pi ^{-1}\Delta \ln
\left\vert \Delta \right\vert , \\
\partial _{\Delta }\zeta _{\mathrm{g}}|_{\Delta \rightarrow 0} &\approx
&-2\pi ^{-1}\ln \left\vert \Delta \right\vert , \\
\partial _{\Delta }^{2}\zeta _{\mathrm{g}}|_{\Delta \rightarrow 0} &\approx
&-2\pi ^{-1}\Delta ^{-1},
\end{eqnarray}%
around the segment $\left\vert \Delta \right\vert =0$\ $(\left\vert \mu
\right\vert \leqslant 1)$.

Similarly, the analytic expressions of $\overline{\zeta }$\ and $\overline{%
\eta }$ help us investigate their analytical behaviors. (i) Around lines $%
\left\vert \mu \right\vert =1$, we have%
\begin{equation}
\partial _{\mu }\overline{\eta }=\pm \frac{1}{2\pi \left\vert \Delta
\right\vert }\ln \frac{4\Delta ^{2}}{\left\vert \mu ^{2}-1\right\vert },
\end{equation}%
which is divergent at the phase boundary $\mu =\pm 1$. By contrast, when $%
\mu \approx \pm 1^{\pm },$ we have 
\begin{equation}
\partial _{\mu }\overline{\zeta }\approx \pm \frac{1}{2\pi \left\vert \Delta
\right\vert }\ln \frac{4\Delta ^{2}}{\left\vert \mu ^{2}-1\right\vert },
\end{equation}%
which is divergent. However, we have $\partial _{\mu }\overline{\zeta }=0$\
at $\left\vert \mu \right\vert =1^{-}$. Therefore, $\partial _{\mu }%
\overline{\zeta }$\ is discontinuous at the lines $\left\vert \mu
\right\vert =1$. (ii) Around the segment $\left\vert \Delta \right\vert =0$ (%
$\left\vert \mu \right\vert <1$),\ we have 
\begin{equation}
\partial _{\Delta }\overline{\eta }\approx \pm 0.5\pi ^{-1}\left[ \ln \left(
\left\vert \left\vert \mu \right\vert +1\right\vert \right) -\ln \left(
\left\vert \left\vert \mu \right\vert -1\right\vert \right) \right]
\end{equation}%
for $\Delta =0^{\pm }$, resulting in the divergent $\partial _{\Delta }^{2}%
\overline{\eta }$. By contrast, we have 
\begin{equation}
\partial _{\Delta }\overline{\zeta }\approx \mp \pi ^{-1}\ln \left(
\left\vert \Delta \right\vert \right)
\end{equation}
for $\Delta =0^{\pm }$. It is divergent at the phase boundary $\left\vert
\Delta \right\vert =0$. In summary, $\partial _{\mu }\overline{\eta },$\ $%
\partial _{\mu }\overline{\zeta }$, $\partial _{\Delta }\overline{\eta },$\
and $\partial _{\Delta }\overline{\zeta }$, can identify the phase boundary
at $\mu =1$\ and $\Delta =0$ $(\mu <1)$. Compared with $\eta _{\mathrm{g}}$
and $\zeta _{\mathrm{g}},$ which are always identical in all regions, $%
\overline{\zeta }$\ and $\overline{\eta }$ vary in non-trivial regions,
providing quantitative description of the pairing mechanism for a
non-equilibrium state. Moreover, the non-equilibrium state is also a
superconducting state because the values of $\overline{\zeta }$\ and $%
\overline{\eta }$ have the same order as those of $\zeta _{\mathrm{g}}$ and $%
\eta _{\mathrm{g}}$. The fact that $\overline{\zeta }>\overline{\eta }$
indicates that BCS-like pairing in momentum space is favored.


\begin{thebibliography}{99}


\bibitem{ZG} G. Zhang and Z. Song, Topological Characterization of Extended
Quantum Ising Models, Phys. Rev. Lett. \textbf{115}, 177204 (2015).

\bibitem{Choi} S. Choi, J. Choi, R. Landig, G. Kucsko, H. Zhou, J. Isoya, F.
Jelezko, S. Onoda, H. Sumiya, V. Khemani, C. v. Keyserlingk, N. Y. Yao, E.
Demler, and M. D. Lukin, Observation of discrete time-crystalline order in a
disordered dipolar many-body system, Nature \textbf{543}, 221 (2017).

\bibitem{Else} D. V. Else, B. Bauer, and C. Nayak, Floquet time crystals,
Phys. Rev. Lett. \textbf{117}, 090402 (2016).

\bibitem{Khemani} V. Khemani, A. Lazarides, R. Moessner, and S. L. Sondhi,
Phase structure of driven quantum systems, Phys. Rev. Lett. \textbf{116},
250401 (2016).

\bibitem{Lindner} N. H. Lindner, G. Refael, and V. Galitski, Floquet
Topological Insulator in Semiconductor Quantum Wells, Nat. Phys. \textbf{7},
490 (2011).

\bibitem{Kaneko} T. Kaneko, T. Shirakawa, S. Sorella, and S. Yunoki,
Photoinduced $\eta $ Pairing in the Hubbard Model, Phys. Rev. Lett. \textbf{%
\ 122}, 077002 (2019).

\bibitem{Tindall} J. Tindall, B. Bu\v{c}a, J.\thinspace R. Coulthard, and D.
Jaksch, Heating-Induced Long-Range \ $\eta $ Pairing in the Hubbard Model,
Phys. Rev. Lett. \textbf{123}, 030603 (2019).

\bibitem{YXMPRA} X. M. Yang and Z. Song, Resonant generation of a p-wave
Cooper pair in a non-Hermitian Kitaev chain at the exceptional point, Phys.
Rev. A \textbf{102}, 022219 (2020).

\bibitem{ZXZPRB2} X. Z. Zhang and Z. Song, $\eta $-pairing ground states in
the non-Hermitian Hubbard model, Phys. Rev. B \textbf{103}, 235153 (2021).

 

\bibitem{TK} T. Kaneko, T. Shirakawa, S. Sorella, and S. Yunoki,
Photoinduced $\eta $ Pairing in the Hubbard Model, Phys. Rev. Lett. \textbf{%
122}, 077002 (2019).

\bibitem{JT1} J. Tindall, F. Schlawin, M. A. Sentef, and D. Jaksch,
Analytical solution for the steady states of the driven Hubbard model, Phys.
Rev. B \textbf{103}, 035146

\bibitem{JT2} J. Tindall, F. Schlawin, M. Sentef and D. Jaksch, Lieb's Theorem and Maximum Entropy Condensates, Quantum \textbf{5}, 610 (2021).

\bibitem{Jochim} {{S. Jochim, M. Bartenstein, A. Altmeyer, G. Hendl, S.
Riedl, C. Chin, }J. Hecker Denschlag, and R. Grimm, Bose-Einstein
Condensation }of Molecules,{\ Science \textbf{302}, 2101 (2003).}

\bibitem{Greiner} {M. Greiner, C. A. Regal, and D. S. Jin, Emergence of a
molecular }Bose--Einstein condensate from a Fermi gas, {Nature (London) 
\textbf{426}, 537 (2003).}



\bibitem{AC} A. Cavalleri, Photo-induced superconductivity, Contemporary
Physics, 59, 31 (2018).

\bibitem{DF} D. Fausti, R. I. Tobey, N. Dean, S. Kaiser, A. Dienst, M. C. Ho
mann, S. Pyon, T. Takayama, H. Takagi, and A. Cavalleri, Light-Induced
Superconductivity in a Stripe-Ordered Cuprate, Science \textbf{331}, 189
(2011).

\bibitem{MM} M. Mitrano, A. Cantaluppi, D. Nicoletti, S. Kaiser, A.
Perucchi, S. Lupi, P. Di Pietro, D. Pontiroli, M. Ricc, S. R. Clark, et al.,
Nature \textbf{530}, 461 (2016).

\bibitem{TS} T. Suzuki, T. Someya, T. Hashimoto, S. Michimae, M. Watanabe,
M. Fujisawa, T. Kanai, N. Ishii, J. Itatani, S. Kasahara, et al.,
Photoinduced possible superconducting state with long-lived disproportionate
band filling in FeSe, Commun. Phys. 2, 115 (2019).

\bibitem{LS} L. Stojchevska, I. Vaskivskyi, T. Mertelj, P. Kusar,
D. Svetin, S. Brazovskii, and D. Mihailovic, Science \textbf{344}, 177
(2014).

\bibitem{HM} H. Matsuzaki, M. Iwata, T. Miyamoto, T. Terashige, K. Iwano, S.
Takaishi, M. Takamura, S. Kumagai, M. Yamashita, R. Takahashi, et al.,
Excitation-Photon-Energy Selectivity of Photoconversions in Halogen-Bridged
Pd-Chain Compounds: Mott Insulator to Metal or Charge-Density-Wave State,
Phys. Rev. Lett. \textbf{113}, 096403 (2014).

\bibitem{AK} A. Kogar, A. Zong, P. E. Dolgirev, X. Shen, J. Straquadine,
Y.-Q. Bie, X. Wang, T. Rohwer, I.-C. Tung, Y. Yang, et al., Light-induced
charge density wave in LaTe3, Nat. Phys. \textbf{16}, 159 (2020).

\bibitem{YM} Y. Murotani, C. Kim, H. Akiyama, L. N. Pfei er, K. W. West, and
R. Shimano, Light-driven electron-hole Bardeen-Cooper-Schrieffer-like state
in bulk GaAs, Phys. Rev. Lett. \textbf{123}, 197401 (2019).

\bibitem{SK} S. Kitamura and H. Aoki, Phys. Rev. B \textbf{94}, 174503
(2016).

\bibitem{JT} J. Tindall, B. Bu ca, J. R. Coulthard, and D. Jaksch, Phys.
Rev. Lett. 123, 030603 (2019).

\bibitem{RF1} R. Fujiuchi, T. Kaneko, Y. Ohta, and S. Yunoki, Phys. Rev. B
100, 045121 (2019).

\bibitem{FP} F. Peronaci, O. Parcollet, and M. Schir o, Phys. Rev. B 101,
161101 (2020).

\bibitem{RF2} R. Fujiuchi, T. Kaneko, K. Sugimoto, S. Yunoki, and Y. Ohta,
Phys. Rev. B 101, 235122 (2020).

\bibitem{SE} S. Ejima, T. Kaneko, F. Lange, S. Yunoki, and H. Fehske, Phys.
Rev. Research \textbf{2}, 032008 (2020).

\bibitem{JL} J. Li, D. Golez, P.Werner, and M. Eckstein, Phys. Rev. B 102,
165136 (2020).

\bibitem{TK2} T. Kaneko, S. Yunoki, and A. J. Millis, Phys. Rev. Research 2,
032027 (2020).

\bibitem{ZXZPRB3} X. Z. Zhang and Z. Song, Steady off-diagonal long-range
order state in a half-filled dimerized Hubbard chain induced by a resonant
pulsed field, Phys. Rev. B. \textbf{106}, 094301(2022).

\bibitem{Kitaev} A. Y. Kitaev, Unpaired Majorana fermions in quantum wires,
Phys. Usp. \textbf{44}, 131 (2001).

\bibitem{Sarma} C. Nayak, S. H. Simon, A. Stern, M. Freedman, and S. D.
Sarma, Non-Abelian anyons and topological quantum computation, Rev. Mod.
Phys. \textbf{80,} 1083 (2008).

\bibitem{Stern} A. Stern, Non-Abelian states of matter, Nature (London) 
\textbf{464,} 187 (2010).

\bibitem{Alicea} J. Alicea, New directions in the pursuit of Majorana
fermions in solid state systems, Rep. Prog. Phys. \textbf{75,} 076501 (2012).



\bibitem{MH} M. Heyl, Dynamical quantum phase transitions: a review, Rep.
Prog. Phys \textbf{81}, 054001(2018).

\bibitem{LZ} L. W. Zhou and Q. Q. Du, Non-Hermitian topological phases and
dynamical quantum phase transitions: a generic connection, New J. Phys. 
\textbf{23}, 063041(2021).

\bibitem{Pfeuty} P. Pfeuty, The one-dimensional Ising model with a
transverse field, Ann. Phys. (NY) \textbf{57}, 79 (1970).

\bibitem{SachdevBook} S. Sachdev, Quantum Phase Transitions (Cambridge
University Press, Cambridge, England, 1999).

\bibitem{TPCHOY} T.-P. Choy, J. M. Edge, A. R. Akhmerov, and C. W. J.
Beenakker, Majorana fermions emerging from magnetic nanoparticles on a
superconductor without spin-orbit coupling, Phys. Rev. B \textbf{84}, 195442
(2011).

\bibitem{AGG} A. Gorczyca-Goraj, T. Domanski, and M. M. Maska, Topological
superconductivity at finite temperatures in proximitized magnetic nanowires,
Phys. Rev. B \textbf{99}, 235430 (2019).

\bibitem{DV1} D. Vodola, L. Lepori, E. Ercolessi, A. V. Gorshkov, and G.
Pupillo, Kitaev Chains with Long-Range Pairing, Phys. Rev. Lett. \textbf{113}%
, 156402 (2014).

\bibitem{DV2} D. Vodola, L. Lepori, E. Ercolessi, A. V. Gorshkov, and G.
Pupillo, Long-range ising and kitaev models: Phases, correlations and edge
modes, New. J. Phys. \textbf{18}, 015001 (2015).

\bibitem{OV} O. Viyuela, D. Vodola, G. Pupillo, and M. A. Martin-Delgado,
Topological massive dirac edge modes and long-range superconducting
hamiltonians, Phys. Rev. B \textbf{94}, 125121 (2016).

\bibitem{LL} L. Lepori and L. Dell'Anna, Long-range topological insulators
and weakened bulk-boundary correspondence, New. J. Phys. \textbf{19}, 103030
(2017).

\bibitem{UB} U. Bhattacharya, S.Maity, A. Dutta, and D. Sen, Critical phase
boundaries of static and periodically kicked long-range kitaev chain, J.
Phys.: Condens. Matter \textbf{31}, 174003 (2019).

\bibitem{CKC} C.-K. Chiu, J. C.\thinspace Y. Teo, A. P. Schnyder, and S.
Ryu, Classification of topological quantum matter with symmetries, Rev. Mod.
Phys. \textbf{88}, 035005 (2016).

\bibitem{NL} N. Leumer, M. Marganska, B. Muralidharan and M. Grifoni, Exact
eigenvectors and eigenvalues of the finite Kitaev chain and its topological
properties, J. Phys.: Condens. Matter \textbf{32} 445502 (2020). 


\bibitem{BP} B. Pahlevanzadeh, P. Sahebsara, David S\'{e}n\'{e}chal, Chiral $%
p$-wave superconductivity in twisted bilayer graphene from dynamical mean
field theory, SciPost Phys. \textbf{11}, 017 (2021).

\bibitem{AS} A. Sen, S. Nandy, K. Sengupta, Entanglement generation in
periodically driven integrable systems: Dynamical phase transitions and
steady state, Phys. Rev. B. \textbf{94}, 214301(2016).

\bibitem{AAM} A. A. Makki, S. Bandyopadhyay, S. Maity, A. Dutta, Dynamical
crossover behavior in the relaxation of quenched quantum many-body systems,
Phys. Rev. B \textbf{105}, 054301 (2022).

\bibitem{SA} S. Aditya, S. Samanta, A. Sen, K. Sengupta, D. Sen, Dynamical
relaxation of correlators in periodically driven integrable quantum systems,
Phys. Rev. B \textbf{105}, 104303 (2022).

\end{thebibliography}
\end{document}